\documentclass[pra,aps,amsmath,amssymb,showkeys]{revtex4}
\usepackage{graphicx}
\newcommand{\hh}{\mathcal{H}}
\newcommand{\pen}{\openone}
\newcommand{\lnp}{\mathcal{L}}
\newcommand{\lsa}{{\mathcal{L}}_{s.a.}}
\newcommand{\lsp}{{\mathcal{L}}_{+}}
\newcommand{\dsp}{{\mathcal{D}}}
\newcommand{\am}{{\mathsf{A}}}
\newcommand{\bn}{{\mathsf{B}}}
\newcommand{\km}{{\mathsf{K}}}
\newcommand{\mm}{{\mathsf{M}}}
\newcommand{\asx}{{\mathsf{S}}}
\newcommand{\ax}{{\mathsf{X}}}
\newcommand{\ay}{{\mathsf{Y}}}

\newcommand{\iu}{{\mathtt{i}}}

\newcommand{\clb}{\mathcal{B}}

\newcommand{\clj}{{\mathcal{J}}}
\newcommand{\clm}{{\mathcal{M}}}
\newcommand{\cln}{{\mathcal{N}}}
\newcommand{\clp}{{\mathcal{P}}}

\newcommand{\bro}{\boldsymbol{\rho}}
\newcommand{\vbro}{\boldsymbol{\varrho}}
\newcommand{\bdl}{\boldsymbol{\xi}}

\newcommand{\bsg}{\boldsymbol{\sigma}}

\newcommand{\wbro}{\widetilde{\boldsymbol{\rho}}}
\newcommand{\wmu}{\widetilde{\mu}}
\newcommand{\wsi}{\widetilde{\psi}}
\newcommand{\wclb}{\widetilde{\mathcal{B}}}
\newcommand{\tr}{{\mathrm{tr}}}
\newcommand{\id}{{\mathrm{id}}}
\newcommand{\rmdd}{{\mathrm{D}}}
\newcommand{\rmc}{{\mathrm{C}}}
\newcommand{\rms}{{\mathrm{S}}}
\newcommand{\rmr}{{\mathrm{R}}}
\newcommand{\rmw}{{\mathrm{W}}}
\newcommand{\spc}{{\mathrm{spec}}}
\newcommand{\ron}{{\mathrm{ran}}}
\newcommand{\dig}{\mathrm{diag}}
\newcommand{\pim}{{\mathsf{\Pi}}}
\newcommand{\zmx}{\mathbf{0}}

\begin{document}
\clearpage
\preprint{}

\title{Coherence quantifiers from the viewpoint of their decreases in the measurement process}

\author{Alexey E. Rastegin}
\affiliation{Department of Theoretical Physics, Irkutsk State University, Russia}

\begin{abstract}
Measurements can be considered as a genuine example of processes that crush quantum coherence. In the case of an observable with degeneracy, the formulations of L\"{u}ders and von Neumann are known. These pictures postulate the two different states of a system immediately following the act of measurement. Hence, they are associated with divers variants of coherence losses during the measurement. Recent studies have focused on several ways to characterize quantum coherence appropriately. One of the existing types of quantifier is based on quantum $\alpha$-divergences of the Tsallis type. In this paper, we introduce coherence quantifiers associated with the L\"{u}ders picture of quantum measurements. The are shown to satisfy the same properties as coherence $\alpha$-quantifiers related to some orthonormal basis. Further, we consider losses of quantum coherence during a generalized measurement. The proposed approach is exemplified with unambiguous state discrimination; extreme properties of the states to be discriminated are clearly shown.
\end{abstract}


\maketitle

\pagenumbering{arabic}
\setcounter{page}{1}

\section{Introduction}\label{sec1}

Theoretical and experimental studies of coherence has a long history
in physics. Complete understanding of this concept could be reached
only within a purely quantum approach. In effect, recent
investigations of coherence are connected with modern prospective
technologies including quantum computations and quantum
cryptography. One of genuine features of coherence-like quantities
is that they are basis dependent. In many physical cases of
interest, only a limited number of bases actually have a priority. This claim is
quite obvious in application to quantum systems of information
processing. The quantum parallelism of Deutsch \cite{deutsch} 
is realized through quantum superpositions written in the prescribed
basis. The concept of the pointer basis plays an important role in
our treatment of measurement process \cite{zurek81}. Thermodynamic
properties of nano-systems at low temperatures are commonly
considered with the use of concrete representation for statistical
mixtures \cite{horodecki15,ngour15}. Contemporary advances in
theoretical studies of quantum correlations are reviewed in
\cite{adesso16jpa,fan2017}.

The characteristics of coherence and decoherence seem to be opposite
to each other. Hence, various coherence quantifiers could be
examined from the viewpoint of their decrease during processes with
deep decoherence. The authors of \cite{yao17} have noted that
quantum measurements are a quite typical example of such processes.
If we adopt, here, the projection postulate, then this consideration
leads us to one of the very core questions of quantum mechanics. The
actual state right after measuring a degenerate observable can be
given in two different forms, due to von Neumann and L\"{u}ders,
respectively. The reduction rule of von Neumann appeals to the fact
that each measurement uses a particular apparatus. Instead of the
degenerate observable {\it per se}, we actually deal with its
refinement (see section V.1 in \cite{neumann32}). The latter
commutes with the former, but has only non-degenerate eigenvalues.
There is an obvious freedom in the choice of such refinements.
L\"{u}ders \cite{luders51} has criticized von Neumann's anzatz and
replaced it with another one. Nowadays, the L\"{u}ders formulation of
the projection postulate is most commonly used.

The relative entropy of coherence and the $\ell_{1}$-norm of
coherence are widely applied due to their useful properties
\cite{bcp14}. The authors of \cite{yao17} extended these quantities
to measurements of the L\"{u}ders type, and mentioned the hierarchy
relations showing a residual coherence. One family of coherence
quantifiers is based on quantum $\alpha$-divergences of the Tsallis
type. In this work, we aim to extend this concept to the case of
L\"{u}ders-type measurements. Together with distance-based
quantifiers of coherence, other quantities deserve to be considered.
In particular, the robustness of coherence \cite{robcoh16} and the
coherence weight \cite{anand17} have recently been proposed. The problem
of maximizing coherence with respect to the reference bases was
addressed in \cite{yao16,hsfan17}. It turned out that bases mutually
unbiased with the state eigenbasis are optimal for the robustness of
coherence and the coherence weight. Generalized quantum measurements
are indispensable in quantum information processing. Basic ways of
quantifying coherence can be extended to measurements described by
positive operator-valued measures (POVMs). We will illustrate these
proposals with unambiguous state discrimination, which is a very
important and intuitively understandable example of a rank-one POVM.

The paper is organized as follows. In section \ref{sec2}, we review
the required material and fix the notation. Some standard results
about quantum operations and measurements will be used throughout the
paper. In particular, we recall both the von Neumann and L\"{u}ders
approaches to measure an observable with degenerate eigenvalues.
Section \ref{sec3} is devoted to coherence quantifiers on the base
of quantum Tsallis $\alpha$-divergences as applicatied to the
L\"{u}ders picture. Basic properties of such quantifiers are
discussed. The so-called residual coherence can be characterized by
means of various coherence measures. Using the example of a concrete
spin observable with a degenerate eigenvalue, we compare the level
of residual coherence predicted by several quantifiers. In section
\ref{sec4}, we address the question how to characterize losses of
quantum coherence during a generalized quantum measurement. In the
case of rank-one POVMs, we propose a natural approach realized
through orthonormal bases in a suitably extended space. This approach
is exemplified with the measurement designed for unambiguous state
discrimination. In section \ref{sec5}, we conclude the paper.

\section{Preliminaries}\label{sec2}

In this section, we begin by recalling the required formal
definitions. Let $\lnp(\hh)$ be the space of linear operators on
finite-dimensional Hilbert space $\hh$. By $\lsp(\hh)$ and
$\lsa(\hh)$, we denote respectively the set of positive
semidefinite operators and the real space of Hermitian ones. A
state of the quantum system of interest is represented by the
density matrix $\bro\in\lsp(\hh)$ normalized as $\tr(\bro)=1$.
Such matrices form the convex set $\dsp(\hh)$ of density operators
acting on $\hh$. The range of $\am\in\lnp(\hh)$ will be denoted as
$\ron(\am)$. For $\am\in\lsp(\hh)$, we define $\am^{0}$ as the
orthogonal projector onto $\ron(\am)$. In finite dimensions, we
treat $\am^{0}\vee\bn^{0}$ as the projector onto the sum of
subspaces $\ron(\am)+\ron(\bn)$. In the infinite-dimensional case,
this definition should be modified. In the following, we will deal
with the finite-dimensional case only. A distance between
operators can be characterized by appropriately chosen norms. With
respect to the given orthonormal basis, each operator
$\am\in\lnp(\hh)$ is represented by the square matrix with
elements $a_{ij}$. The $\ell_{1}$-norm is then defined as
\cite{hornJ}
\begin{equation}
\|\am\|_{\ell_{1}}:=\sum\nolimits_{ij} |a_{ij}|
\, . \label{ell1n}
\end{equation}
There are many norms that can be used to define measures of
distinguishability of quantum states \cite{watrous1}. The
well-known norm (\ref{ell1n}) gives the so-called $\ell_{1}$-norm
of coherence \cite{bcp14}.

Another approach to compare quantum states is based on the notion
of quantum relative entropy, or divergence. This concept is
fundamental in quantum information theory \cite{nielsen,vedral02}.
For $\bro,\vbro\in\dsp(\hh)$, the relative entropy of $\bro$ with
respect to $\vbro$ is written as \cite{hmpb11}
\begin{equation}
\rmdd_{1}(\bro||\vbro):=
\begin{cases}
\tr(\bro\ln\bro-\bro\ln\vbro) \,,
& \text{if $\ron(\bro)\subseteq\ron(\vbro)$} \, , \\
+\infty\, , & \text{otherwise} \, .
\end{cases}
\label{relab}
\end{equation}
It is a quantum counterpart of the standard relative entropy of
probability distributions. For the given probability distributions
$\{p_{j}\}$ and $\{q_{j}\}$, it is defined by \cite{nielsen}
\begin{equation}
D_{1}(p_{j}||q_{j}):=
\sum\nolimits_{j} p_{j}\,\ln\frac{p_{j}}{q_{j}}
\ . \label{relen1}
\end{equation}
If there exists some $j$ such that $p_{j}\neq0$ and $q_{j}=0$,
then the right-hand side of (\ref{relen1}) is set up to be
$+\infty$. General properties of the relative entropies and other
entropic functions are discussed in \cite{nielsen,bengtsson}.

Several generalizations of the above quantities have found
use in various topics \cite{icsr08}. For $0<\alpha\neq1$, the
Tsallis relative $\alpha$-entropy is defined as
\cite{borland,sf04}
\begin{equation}
D_{\alpha}(p_{j}||q_{j}):=
\frac{1}{\alpha-1}
\left(
\sum\nolimits_{j} p_{j}^{\alpha}q_{j}^{1-\alpha}-1
\right)
. \label{tfdf}
\end{equation}
If for some $j$ we have $p_{j}\neq0$ and $q_{j}=0$ simultaneously,
then the relative $\alpha$-entropy with $\alpha>1$ is taken as
$+\infty$. In the limit $\alpha\to1$, the quantity (\ref{tfdf})
gives the standard relative entropy (\ref{relen1}). The formula
(\ref{tfdf}) can be represented similarly to (\ref{relen1}) with
the use of the $\alpha$-logarithm. It is easy to see that
$D_{\alpha}(p_{j}||q_{j})\geq0$. Necessary conditions for
vanishing $D_{\alpha}(p_{j}||q_{j})$ follow from the results of
\cite{vajda06}. Using example 2 of \cite{vajda06}, we can prove
that $D_{\alpha}(p_{j}||q_{j})=0$ only if $p_{j}=q_{j}$ for all
$j$. The relative $\alpha$-entropy (\ref{tfdf}) is a particular
case of the Csisz\'{a}r $f$-divergences \cite{ics67}.

Quantum $f$-divergences were examined in detail in \cite{hmpb11}.
This approach allows us to involve relative $\alpha$-entropies of
the Tsallis type. It will be useful to define them for arbitrary
positive semidefinite operators. Let $\am$ and $\bn$ be positive
operators such that $\ron(\am)\subseteq\ron(\bn)$. For
$0<\alpha\neq1$, the Tsallis $\alpha$-divergence of $\am$ with
respect to $\bn$ is defined as
\begin{equation}
\rmdd_{\alpha}(\am||\bn):=
\frac{1}{\alpha-1}\Bigl[\tr(\am^{\alpha}\bn^{1-\alpha})-\tr(\am)\Bigr]
. \label{dfab}
\end{equation}
Since $\ron(\am)\subseteq\ron(\bn)$, the trace should be taken
over $\ron(\bn)$. For $\alpha\in(0;1)$, the expression
(\ref{dfab}) is used without such conditions. Several properties
of the quantum $\alpha$-divergence follow from the corresponding
results on the quantum $f$-divergences \cite{hmpb11}. For all
$\lambda\in[0;+\infty)$, one satisfies
\begin{equation}
\rmdd_{\alpha}(\lambda\am||\lambda\bn)=\lambda\,\rmdd_{\alpha}(\am||\bn)
\, . \label{flam}
\end{equation}
Let four positive semidefinite operators $\am_{1}$, $\bn_{1}$,
$\am_{2}$, $\bn_{2}$ obey
$\am_{1}^{0}\vee\bn_{1}^{0}\perp\am_{2}^{0}\vee\bn_{2}^{0}$; then
\begin{equation}
\rmdd_{\alpha}\bigl(\am_{1}+\am_{2}\big|\big|\bn_{1}+\bn_{2}\bigr)
=\rmdd_{\alpha}(\am_{1}||\bn_{1})+\rmdd_{\alpha}(\am_{2}||\bn_{2})
\, . \label{tdit}
\end{equation}
The latter can be proved for quantum $f$-divergences under certain
conditions \cite{hmpb11}.

One of fundamental properties of the quantum relative entropy is
its monotonicity under trace-preserving completely positive maps
\cite{nielsen}. In the classical regime, the relative Tsallis
entropy (\ref{tfdf}) is monotone under stochastic maps for all
$\alpha\geq0$ \cite{sf04}. This is not the case for the quantum
regime. Let us recall basic facts about quantum operations. We
consider a linear map
\begin{equation}
\Phi:\>\lnp(\hh)\rightarrow\lnp(\hh^{\prime})
\, , \label{hdhdp}
\end{equation}
where the input space $\hh$ and the output space $\hh^{\prime}$
may differ. This map is positive, when
$\Phi(\am)\in\lsp(\hh^{\prime})$ for each $\am\in\lsp(\hh)$
\cite{nielsen}. Physical processes are described by completely
positive maps \cite{nielsen}. Let $\id^{\prime\prime}$ be the
identity map on $\lnp(\hh^{\prime\prime})$, where the Hilbert
space $\hh^{\prime\prime}$ is related to an imagined reference
system. The complete positivity implies that the map
$\Phi\otimes\id^{\prime\prime}$ is positive for arbitrary
dimensionality of $\hh^{\prime\prime}$. Each completely positive
map can be represented in the form \cite{nielsen,watrous1}
\begin{equation}
\Phi(\am)=\sum\nolimits_{i}\km_{i}\am\km_{i}^{\dagger}
\, , \label{osrp}
\end{equation}
with the Kraus operators $\km_{i}:{\>}\hh\rightarrow\hh^{\prime}$.
The map preserves the trace, when these operators obey
\begin{equation}
\sum\nolimits_{i}\km_{i}^{\dagger}\km_{i}=\pen
\, , \label{clrl}
\end{equation}
where $\pen$ denotes the identity on $\hh$. Trace-preserving
completely positive (TPCP) maps are usually referred to as quantum
channels \cite{nielsen}.

The quantum $\alpha$-divergence is monotone under TPCP maps for
$\alpha\in(0;2]$, so that
\begin{equation}
\rmdd_{\alpha}\bigl(\Phi(\bro)\big|\big|\Phi(\vbro)\bigr)\leq\rmdd_{\alpha}(\bro||\vbro)
\, . \label{mnren}
\end{equation}
This inequality follows from theorem 4.3 of \cite{hmpb11} together
with some facts about functions on positive matrices. The
monotonicity also implies the joint convexity of the
$f$-divergences in line with corollary 4.7 of \cite{hmpb11}. In
particular, the quantum $\alpha$-divergences of the Tsallis type
are jointly convex for $\alpha\in(0;2]$. Let $\{\bro_{i}\}$ and
$\{\vbro_{i}\}$ be two collections of density matrices, and let
$q_{i}$'s be positive numbers that sum to $1$. For
$\alpha\in(0;2]$, we then have
\begin{equation}
\rmdd_{\alpha}\biggl(\sum\nolimits_{i}q_{i}\bro_{i}\bigg|\bigg|\sum\nolimits_{i}q_{i}\vbro_{i}\biggr)
\leq\sum\nolimits_{i}q_{i}\,\rmdd_{\alpha}(\bro_{i}||\vbro_{i})
\, . \label{joico}
\end{equation}
The properties (\ref{mnren}) and (\ref{joico}) are important in
the verification of corresponding properties of induced coherence
measures.

The description of quantum measurements is indispensable in the
sense that without it the quantum-mechanical formalism is not
complete. Let us consider some observable $\ax\in\lsa(\hh)$ with
the spectral decomposition
\begin{equation}
\ax=\sum\nolimits_{j} x_{j}\pim_{j}
\, . \label{spdx}
\end{equation}
In this sum, the eigenvalue labels $x_{j}\in\spc(\ax)$ are all
assumed to be different. For the pre-measurement state $\bro$, the
$j$th outcome occurs with the probability $\tr(\pim_{j}\bro)$.
Another question to be resolved concerns the form of the state immediately
following the act of measurement. Any answer to this question is actually a
kind of reduction rule. In the following, we focus on measurements
that obey the projection postulate. In this case, there are two
different ways to treat quantum measurements of an observable with
degenerate eigenvalues. Then the Hilbert space $\hh$ is
correspondingly represented as the direct sum
\begin{equation}
\hh=\bigoplus\nolimits_{j}\hh_{j}
\, , \qquad
\hh_{j}=\ron(\pim_{j})
\, , \label{hdsn}
\end{equation}
so that $|\psi\rangle\in\hh_{k}$ implies
$\pim_{j}|\psi\rangle=\delta_{kj}|\psi\rangle$ for all $j$. The
two answers to the question are respectively due to von Neumann
\cite{neumann32} and L\"{u}ders \cite{luders51}. We begin with the
latter, since now it is commonly accepted by the community.

Suppose that the pre-measurement state is described by density
matrix $\bro$. The so-called L\"{u}ders rule claims that the
post-measurement state is represented by
\begin{equation}
\Phi_{\clp}(\bro)=\sum\nolimits_{j} \pim_{j}\bro\,\pim_{j}
\, . \label{lurr}
\end{equation}
Here, we actually deal with TPCP map
$\Phi_{\clp}:\>\lnp(\hh)\rightarrow\lnp(\hh)$ assigned to the set
$\clp=\bigl\{\pim_{j}\bigr\}$ of operators of orthogonal
projection. According to (\ref{lurr}), we introduce the set of
invariant states:
\begin{equation}
\clj_{\clp}:=\Bigl\{\bdl:\>\bdl\in\dsp(\hh),\>\Phi_{\clp}(\bdl)=\bdl\Bigr\}
\, . \label{invlud}
\end{equation}
This definition is similar to the definition of the set of
symmetric states in resources theories of asymmetry
\cite{mspekk13,robass16}. In the following, the set (\ref{invlud})
of invariant states will be applied to define coherence
quantifiers associated with the L\"{u}ders reduction rule.

The first complete treatment of the measurement problem was given by
von Neumann \cite{neumann32}. His reduction rule is slightly more
complicated to formulate. Instead of $\ax$, we should considers some
its refinement $\ay$. The latter commutes with $\ax$ but has only
non-degenerate eigenvalues. In this way, we obtain an indirect
measurement of the observable to be measured. A concrete example of
indirect spin measurement is described in \cite{mayato12}. The problem of discriminating measurement contexts was
analyzed in general in \cite{mayato11}. The authors of \cite{mayato11} also
noted that their results allow one to check experimentally
whether an apparatus performs a L\"{u}ders or a von Neumann
measurement. This proposal was successfully implemented in
\cite{shukla16}. The spectral decomposition of $\ay$ can be
expressed as
\begin{equation}
\ay=\sum\nolimits_{j\beta} y_{j\beta}
\,|y_{j\beta}\rangle\langle{y}_{j\beta}|
\, , \label{spdxp}
\end{equation}
so that each subspace $\hh_{j}$ is spanned by the vectors
$|y_{j\beta}\rangle$. There exists a function $y\mapsto{g}(y)$
with the following property. For each $x_{j}\in\spc(\ax)$, the
equality $g(y_{j\beta})=x_{j}$ takes place for all $\beta$. The
von Neumann rule actually refers to the orthonormal basis
$\clb=\bigl\{|y_{j\beta}\rangle\bigr\}$. This rule then postulates
the post-measurement state
\begin{equation}
\Phi_{\clb}(\bro)=\sum\nolimits_{j\beta}
|y_{j\beta}\rangle\langle{y}_{j\beta}|\bro|y_{j\beta}\rangle\langle{y}_{j\beta}|
\, . \label{vnrr}
\end{equation}
Hence, the corresponding set of invariant states reads as
\begin{equation}
\clj_{\clb}:=\Bigl\{\bdl:\>\bdl\in\dsp(\hh),\>\Phi_{\clb}(\bdl)=\bdl\Bigr\}
\, . \label{invlvn}
\end{equation}
The set (\ref{invlvn}) contains all the states that are incoherent
with respect to the basis $\clb$. As a refinement $\ay$ of $\ax$
is not uniquely defined, we actually deal with a family of sets of
the form (\ref{invlvn}). In the case of observables without
degeneracy, the two forms of the reduction rule discussed above
coincide.

Both the above pictures deal with projective measurements. At the
same time, measurements of more general type are widely used in
quantum information science. Such measurements are described by
positive operator-valued measures. Let $\clm=\{\mm_{j}\}$ be a set
of elements of $\lsp(\hh)$, satisfying the completeness relation
\begin{equation}
\sum\nolimits_{j=1}^{N}\mm_{j}=\pen
\, . \label{comrn}
\end{equation}
Such operators form a POVM. For
the pre-measurement state $\bro$, the probability of $j$th
outcome is written as $\tr(\mm_{j}\bro)$. In contrast to
projective measurements, the number $N$ of different outcomes in a
POVM-measurement can exceed $d={\mathrm{dim}}(\hh)$. In many
tasks, the optimal POVM can be built of rank-one elements
\cite{davies78}. In the following, we will consider coherence
losses in measurements described by rank-one POVMs.

\section{Coherence quantifiers for the L\"{u}ders-type measurements}\label{sec3}

In this section, we will examine properties of some coherence
quantifiers associated with the L\"{u}ders picture. The authors of
\cite{yao17} considered this question with respect to the
$\ell_{1}$-norm of coherence and the relative entropy of coherence.
Initially, measures of quantum coherence with respect to a concrete
orthonormal basis were examined in \cite{bcp14}. In the context of
resource theories, the problem of quantifying coherence is reviewed
in \cite{mspekk16,chitam2016,plenio16}. The $\ell_{1}$-norm of
coherence and the relative entropy of coherence are respectively
introduced as
\begin{align}
\rmc_{\ell_{1}}^{(\clb)}(\bro)
&:=\min\bigl\{\|\bro-\bdl\|_{\ell_{1}}:\>\bdl\in\clj_{\clb}\bigr\}
\, , \label{cel1b}\\
\rmc_{1}^{(\clb)}(\bro)
&:=\min\bigl\{\rmdd_{1}(\bro||\bdl):\>\bdl\in\clj_{\clb}\bigr\}
\, , \label{cre1b}
\end{align}
where $\clj_{\clb}$ is specified by (\ref{invlvn}). These
quantities are both basis dependent. There are well known
expressions for them, viz.
\begin{align}
\rmc_{\ell_{1}}^{(\clb)}(\bro)
&=\sum_{k\gamma\neq{j}\beta}
\bigl|\langle{y}_{k\gamma}|\bro|y_{j\beta}\rangle\bigr|
 , \label{cel1be}\\
\rmc_{1}^{(\clb)}(\bro)
&=H_{1}\bigl(p_{j\beta}\bigr)-\rms_{1}(\bro)
\, . \label{cre1be}
\end{align}
Here, $p_{j\beta}=\langle{y}_{j\beta}|\bro|y_{j\beta}\rangle$ is
the corresponding probability,
$H_{1}\bigl(p_{j\beta}\bigr)=-\sum_{j\beta}p_{j\beta}\ln{p}_{j\beta}$
is the Shannon entropy, and $\rms_{1}(\bro)=-\,\tr(\bro\ln\bro)$ is
the von Neumann entropy of $\bro$. For the von Neumann reduction
rule, we should fix the chosen refinement of an observable with
degenerate eigenvalues. The $\ell_{1}$-norm of coherence and the
relative entropy of coherence seem to be very widely used
measures. Using the $\ell_{1}$-norm of coherence, duality
relations between the coherence and path information were examined
in \cite{bera15,bagan16,qureshi17}. An operational interpretation
of the $\ell_{1}$-norm of coherence was proposed in \cite{apwl17}.
The relative entropy of coherence is useful in formulating
complementarity \cite{hall15,pzflf16} and uncertainty relations
for quantum coherence \cite{pati16,baietal6,rastf18}.

Taking the L\"{u}ders rule, the authors of \cite{yao17} have
proposed the following extensions of (\ref{cel1b}) and
(\ref{cre1b}). In our notation, the corresponding quantities are
represented as
\begin{align}
\rmc_{\ell_{1}}^{(\clp)}(\bro)
&:=\min\bigl\{\|\bro-\bdl\|_{\ell_{1}}:\>\bdl\in\clj_{\clp}\bigr\}
\, , \label{cel1p}\\
\rmc_{1}^{(\clp)}(\bro)
&:=\min\bigl\{\rmdd_{1}(\bro||\bdl):\>\bdl\in\clj_{\clp}\bigr\}
\, , \label{cre1p}
\end{align}
where $\clj_{\clp}$ is formally posed by (\ref{invlud}). Simple
calculations finally result in the formula
\begin{equation}
\rmc_{1}^{(\clp)}(\bro)
=H_{1}(p_{j})-\rms_{1}(\bro)
\, , \label{cre1pp}
\end{equation}
where $p_{j}=\tr(\pim_{j}\bro)$. The right-hand side of (\ref{cre1pp})
does not depend on refinements of $\ax$. It can also be shown that
(\ref{cel1p}) is expressed as \cite{yao17}
\begin{equation}
\rmc_{\ell_{1}}^{(\clp)}(\bro)=\sum\nolimits_{k\neq{j}}
\bigl\|\pim_{k}\bro\,\pim_{j}\bigr\|_{\ell_{1}}
\, . \label{cel1pp}
\end{equation}
Since the definition (\ref{ell1n}) is basis dependent, the
quantifier (\ref{cel1pp}) generally depends not only on the set
$\clp$ of projectors. It is not mentioned explicitly, but the
right-hand side of (\ref{cel1pp}) is also referred to the taken
basis $\clb$. In this sense, the definition depends on the chosen
refinement as well.

Let us proceed to the quantities based on the Tsallis relative
$\alpha$-entropies. With respect to an orthonormal basis, such
quantities were proposed in \cite{rastpra16}. For $\alpha>0$, one
defines
\begin{equation}
\rmc_{\alpha}^{(\clb)}(\bro):=
\min\bigl\{\rmdd_{\alpha}(\bro||\bdl):\>\bdl\in\clj_{\clb}\bigr\}
\, . \label{cabdt}
\end{equation}
Of course, this definition is related to the von Neumann rule. For
the L\"{u}ders case, the corresponding $\alpha$-quantifier is
similarly expressed as
\begin{equation}
\rmc_{\alpha}^{(\clp)}(\bro):=
\min\bigl\{\rmdd_{\alpha}(\bro||\bdl):\>\bdl\in\clj_{\clp}\bigr\}
\, . \label{capdt}
\end{equation}
It immediately follows that $\rmc_{\alpha}^{(\clp)}(\bro)\geq0$
with equality if and only if $\bro\in\clj_{\clp}$. This conclusion
reflects the fact that $\rmdd_{\alpha}(\bro||\vbro)=0$ is
equivalent to $\bro=\vbro$. The optimization problem (\ref{capdt})
can be treated in line with reasons given in \cite{rastpra16}. The
following statement takes place.

\newtheorem{prop41}{Theorem}
\begin{prop41}\label{thm41}
For all $0<\alpha\neq1$, the coherence $\alpha$-quantifier is
expressed by
\begin{equation}
\rmc_{\alpha}^{(\clp)}(\bro)=
\frac{1}{\alpha-1}
\left\{
\Bigl(
\sum\nolimits_{j}\tr(\pim_{j}\bro^{\alpha})^{1/\alpha}
\Bigr)^{\!\alpha}
-1
\right\}
. \label{tres1}
\end{equation}
\end{prop41}

{\bf Proof.} We will assume that $\alpha\neq1$. As the
$\alpha$-divergence $\rmdd_{\alpha}(\bro||\bdl)$ should be
minimized, we further assume $\ron(\bro)\subseteq\ron(\bdl)$. In
the spectral decomposition
\begin{equation}
\bdl=\sum\nolimits_{j}\xi_{j}\pim_{j}
\, , \label{bdlspec}
\end{equation}
we set up $\xi_{j}=0$ whenever $\tr(\pim_{j}\bro)=0$. Due to
(\ref{bdlspec}), we can write
\begin{equation}
\rmdd_{\alpha}(\bro||\bdl)=\frac{1}{\alpha-1}
\left\{
\sum\nolimits_{j}\xi_{j}^{1-\alpha}\,\tr(\pim_{j}\bro^{\alpha})
-1\right\}
, \label{renp1}
\end{equation}
where the sum is taken over non-zero $\xi_{j}$ values. We now
introduce the probabilities $b_{j}$ such that
$b_{j}^{\alpha}\propto\tr\bigl(\pim_{j}\bro^{\alpha}\bigr)$.
Together with the normalization condition, the latter gives
\begin{align}
b_{j}&=
\frac{\tr(\pim_{j}\bro^{\alpha})^{1/\alpha}}{\cln}
\ , \label{qnt1}\\
\cln&=\sum\nolimits_{j}\tr(\pim_{j}\bro^{\alpha})^{1/\alpha}
. \label{qnt2}
\end{align}
Thus, the probabilities (\ref{qnt1}) are uniquely defined for the
prescribed $\bro$ and $\alpha$. Combining
$\tr(\pim_{j}\bro^{\alpha})=\cln^{\alpha}b_{j}^{\alpha}$ with
(\ref{renp1}), one gets
\begin{equation}
\rmdd_{\alpha}(\bro||\bdl)=
\cln^{\alpha}D_{\alpha}(b_{j}||\xi_{j})
+\frac{\cln^{\alpha}-1}{\alpha-1}
\ . \label{renp3}
\end{equation}
Here, the probabilities $b_{j}$ and the denominator $\cln$ depend
on $\bro$ and $\alpha$. So, the variables $\xi_{j}$ take place
only in the first term of the right-hand side of (\ref{renp3}).
Since $D_{\alpha}(b_{j}||\xi_{j})\geq0$, the minimal value of
(\ref{renp3}) is reached by setting $\xi_{j}=b_{j}$ with
$D_{\alpha}(b_{j}||\xi_{j})=0$. The corresponding state is
expressed as
\begin{equation}
\bdl^{\star}=\sum\nolimits_{j}b_{j}\pim_{j}
\, . \label{bdlstar}
\end{equation}
Combining this with (\ref{qnt2}) leads to the right-hand side of
(\ref{tres1}). $\blacksquare$

It is important that the quantifier (\ref{capdt}) is convex for
$\alpha\in(0;2]$. We can derive this conclusion from
(\ref{joico}). Let $\{\bro_{i}\}$ be a collection of density
matrices, and let positive numbers $q_{i}$ obey $\sum_{i}q_{i}=1$.
For all $\alpha\in(0;2]$, we have
\begin{equation}
\rmc_{\alpha}\Bigl(\sum\nolimits_{i}q_{i}\bro_{i}\Bigr)
\leq\sum\nolimits_{i}q_{i}\,\rmc_{\alpha}(\bro_{i})
\, . \label{conmixa}
\end{equation}
Let $\Upsilon:\>\lnp(\hh)\rightarrow\lnp(\hh)$ be a TPCP map that
leaves the set $\clj_{\clp}$ to be invariant. For
$\alpha\in(0;2]$, the coherence quantifier (\ref{capdt}) is
monotone under this quantum operation, so that
\begin{equation}
\rmc_{\alpha}\bigl(\Upsilon(\bro)\bigr)\leq\rmc_{\alpha}(\bro)
\, .  \label{crp2}
\end{equation}
The latter follows from the property (\ref{mnren}) and the
definition (\ref{capdt}), which includes the minimization.
Monotonicity under incoherent selective measurements is more
sophisticated \cite{bcp14}. Extending the approach of
\cite{rastpra16}, we pose the monotonicity property as follows.

\newtheorem{prop42}[prop41]{Theorem}
\begin{prop42}\label{thm42}
Let Kraus
operators of TPCP map $\Upsilon:{\>}\lnp(\hh)\rightarrow\lnp(\hh)$
obey the property
\begin{equation}
\km_{i}\clj_{\clp}\km_{i}^{\dagger}\subseteq\clj_{\clp}
\, . \label{krasup}
\end{equation}
For all $\alpha\in(0;2]$, coherence quantifiers of the form
(\ref{capdt}) satisfy
\begin{equation}
\sum\nolimits_{i}q_{i}^{\alpha}s_{i}^{1-\alpha}\,\rmc_{\alpha}(\bro_{i})
\leq\rmc_{\alpha}(\bro)
\, , \label{insla}
\end{equation}
where $q_{i}=\tr(\km_{i}\bro\km_{i}^{\dagger})$,
$\bro_{i}=q_{i}^{-1}\km_{i}\bro\km_{i}^{\dagger}$, and the
probabilities $s_{i}=\tr(\km_{i}\bdl^{\star}\km_{i}^{\dagger})$
are calculated with the state (\ref{bdlstar}).
\end{prop42}

{\bf Proof.} The output of the quantum channel $\Upsilon$ is
represented as
\begin{equation}
\Upsilon(\bro)=\sum\nolimits_{i} q_{i}\bro_{i}
\, . \label{psiib1}
\end{equation}
In terms of the particular outputs
$\bdl_{i}^{\star}=s_{i}^{-1}\km_{i}\bdl^{\star}\km_{i}^{\dagger}$,
we have
\begin{align}
\rmdd_{\alpha}(\bro||\bdl^{\star})
&\geq\sum\nolimits_{i}\rmdd_{\alpha}
\bigl(\km_{i}\bro\km_{i}^{\dagger}
\big|\big|\km_{i}\bdl^{\star}\km_{i}^{\dagger}\bigr)
\label{cmon2}\\
&\geq
\sum\nolimits_{i}q_{i}^{\alpha}s_{i}^{1-\alpha}\,
\rmdd_{\alpha}(\bro_{i}||\bdl_{i}^{\star})
\, . \label{cmon4}
\end{align}
Here, the step (\ref{cmon2}) follows from (\ref{tdit}), and the
step (\ref{cmon4}) follows from theorem 2 of \cite{rastpra16}.
Combining (\ref{cmon4}) with (\ref{capdt}) finally gives
(\ref{insla}). $\blacksquare$

Similarly to (\ref{cabdt}), the quantifier (\ref{capdt}) obeys the
generalized form of monotonicity. For $\alpha=1$, this form is
reduced to the regular form. Hence, for $\alpha\in(0;2]$ the
coherence quantifier (\ref{capdt}) can be treated as a measure with
all required properties. Overall, coherence $\alpha$-quantifiers
associated with the L\"{u}ders picture satisfy the same properties
as coherence $\alpha$-quantifiers related to some orthonormal basis.
It is natural that they succeed only the generalized form of
monotonicity. When $\alpha=1$, the left-hand side of (\ref{insla})
can be interpreted as an averaged output coherence. This view is
somehow similar to the relation
$\rmc_{1}\bigl(\Upsilon(\bro)\bigr)\leq\rmc_{1}(\bro)$. The case $\alpha\neq1$
is more sophisticated, since averaging deals here with weights
$\omega_{i}$ such that
$\bigl[1+(\alpha-1)D_{\alpha}(q_{j}||s_{j})\bigr]\,\omega_{i}=q_{i}^{\alpha}s_{i}^{1-\alpha}$.
Then the left-hand side of (\ref{insla}) is written as the weighted
average of output $\alpha$-quantifiers multiplied by an additional
factor. It provides an interrelation between the relative
$\alpha$-entropy $D_{\alpha}(q_{j}||s_{j})$ and coherence
$\alpha$-quantifiers at the input and output. This relation may be
used when two of three components can be calculated or evaluated, at
least for some $\alpha$.

Let us address the robustness of coherence and the coherence weight.
To each invariant set of states, we assign measures of how far is
the given state from this set. The robustness of asymmetry was
proposed as a measure of asymmetry of quantum states with many
attractive properties \cite{robcoh16,robass16}. The robustness of
coherence is naturally obtained, when we refer to the set of states
diagonal in the prescribed basis. This measure quantifies the
minimal mixing required to destroy all the coherence in a quantum
state \cite{robass16}. In our notation, we have
\begin{equation}
\rmr^{(\clb)}(\bro):=
\min\!\left\{
r\geq0:\>\vbro\in\dsp(\hh),\>\frac{\bro+r\vbro}{1+r}=:\bdl\in\clj_{\clb}\right\}
 . \label{rocbp}
\end{equation}
In this way, one characterizes a coherence change with respect to
the von Neumann rule. For the L\"{u}ders rule, the above term
should be reformulated. Specifically, we put the quantity
\begin{equation}
\rmr^{(\clp)}(\bro):=
\min\!\left\{
r\geq0:\>\vbro\in\dsp(\hh),\>\frac{\bro+r\vbro}{1+r}=:\bdl\in\clj_{\clp}\right\}
 . \label{rocp}
\end{equation}
Let us discuss basic properties of the new quantifier
(\ref{rocp}). As directly follows from this definition, the
equality $\rmr^{(\clp)}(\bro)=0$ is equivalent to
$\bro\in\clj_{\clp}$. Convexity is one of nice properties of the
measure (\ref{rocbp}) and remains valid for (\ref{rocp}), that is
\begin{equation}
\rmr^{(\clp)}\bigl(t\bro_{1}+(1-t)\bro_{2}\bigr)
\leq{t}\,\rmr^{(\clp)}(\bro_{1})+(1-t)\,\rmr^{(\clp)}(\bro_{2})
\, , \label{rconv}
\end{equation}
where $\bro_{1},\bro_{2}\in\dsp(\hh)$ and $t\in[0;1]$. To justify
(\ref{rconv}), we appropriately recast the proof of convexity of
(\ref{rocbp}). Further, we consider a TPCP map
$\Upsilon:\>\lnp(\hh)\rightarrow\lnp(\hh)$ with Kraus operators that
all obey (\ref{krasup}). The quantity (\ref{rocp}) cannot increase
under the action of such operations, i.e.
\begin{equation}
\sum\nolimits_{i} q_{i}\,\rmr^{(\clp)}(\bro_{i})
\leq\rmr^{(\clp)}(\bro)
\, , \label{monro}
\end{equation}
where $q_{i}=\tr(\km_{i}\bro\km_{i}^{\dagger})$ and
$\bro_{i}=q_{i}^{-1}\km_{i}\bro\km_{i}^{\dagger}$. Again, we could
repeat the reasons given in \cite{robcoh16} for the
measure (\ref{rocbp}). We refrain from presenting the details
here.

The authors of \cite{anand17} have proposed the concept of asymmetry
and coherence weight of quantum states. Using the orthonormal basis 
$\clb$, the coherence weight is defined as
\begin{equation}
\rmw^{(\clb)}(\bro):=
\min\Bigl\{
w\geq0:\>\bdl\in\clj_{\clb},\>\vbro\in\dsp(\hh),\>\bro=(1-w)\bdl+w\vbro\Bigr\}
\, . \label{wecbp}
\end{equation}
This measure will be used to characterize coherence changes
according to the von Neumann rule. In a similar manner, we further
write
\begin{equation}
\rmw^{(\clp)}(\bro):=
\min\Bigl\{
w\geq0:\>\bdl\in\clj_{\clp},\>\vbro\in\dsp(\hh),\>\bro=(1-w)\bdl+w\vbro\Bigr\}
\, . \label{wecp}
\end{equation}
The latter is related to (\ref{wecbp}) just as the quantifier
(\ref{rocp}) is related to (\ref{rocbp}). Concerning (\ref{wecp}),
we first note that $W^{(\clp)}(\bro)=0$ is equivalent to
$\bro\in\clj_{\clp}$. As was shown in \cite{anand17}, the quantity
(\ref{wecbp}) is convex as well. The new quantifier (\ref{wecp})
possesses this useful property, i.e.
\begin{equation}
\rmw^{(\clp)}\bigl(t\bro_{1}+(1-t)\bro_{2}\bigr)
\leq{t}\,\rmw^{(\clp)}(\bro_{1})+(1-t)\,\rmw^{(\clp)}(\bro_{2})
\, , \label{wconv}
\end{equation}
where $\bro_{1},\bro_{2}\in\dsp(\hh)$ and $t\in[0;1]$. Further, the
quantity (\ref{wecp}) is monotone under incoherent operations.
If Kraus operators of the quantum channel
$\Upsilon:{\>}\lnp(\hh)\rightarrow\lnp(\hh)$ all obey
(\ref{krasup}), then
\begin{equation}
\sum\nolimits_{i} q_{i}\,\rmw^{(\clp)}(\bro_{i})
\leq\rmw^{(\clp)}(\bro)
\, , \label{monwe}
\end{equation}
where $q_{i}=\tr(\km_{i}\bro\km_{i}^{\dagger})$ and
$\bro_{i}=q_{i}^{-1}\km_{i}\bro\km_{i}^{\dagger}$. We could prove
(\ref{wconv}) and (\ref{monwe}) by adopting the reasons given in
\cite{anand17} for the quantity (\ref{wecbp}).

Comparing coherence quantifiers in the L\"{u}ders and von Neumann
pictures, we at once note the following important fact. For an
observable with degeneracy, one clearly has
$\clj^{(\clb)}\subset\clj^{(\clp)}$. For all the considered ways
to quantify coherence, the minimization is taken under more
conditions in the case of orthonormal bases. Hence, we obtain
\begin{equation}
\rmc^{(\clb)}(\bro)\geq\rmc^{(\clp)}(\bro)
\, , \label{hierr}
\end{equation}
where $\rmc$ can be substituted with the $\ell_{1}$-norm of
coherence, the coherence $\alpha$-quantifier, the robustness of
coherence, and the coherence weight. The authors of \cite{yao17}
mentioned (\ref{hierr}) for the $\ell_{1}$-norm and the relative
entropy of coherence. We only note that the result
(\ref{hierr}) holds in more general context.

We shall now proceed to the following question. Let $\bro$ be the
state right before measurement of an observable $\ax$ with
degenerate spectrum. The post-measurement state can be taken
either as $\Phi_{\clb}(\bro)\in\clj_{\clb}$ due to the von Neumann
rule or as $\Phi_{\clp}(\bro)\in\clj_{\clp}$ due to the L\"{u}ders
rule. Decrease of the amount of coherence can be
characterized by the differences
\begin{align}
\rmc^{(\clb)}(\bro)-\rmc^{(\clb)}\bigl(\Phi_{\clb}(\bro)\bigr)&=\rmc^{(\clb)}(\bro)
\, , \label{dcwrb}\\
\rmc^{(\clp)}(\bro)-\rmc^{(\clp)}\bigl(\Phi_{\clp}(\bro)\bigr)&=\rmc^{(\clp)}(\bro)
\, , \label{dcwrp}
\end{align}
where $\rmc^{(\clb)}$ and $\rmc^{(\clp)}$ are the chosen
quantifiers. In this sense, the quantity
$\Delta\rmc(\bro):=\rmc^{(\clb)}(\bro)-\rmc^{(\clp)}(\bro)$
describes distinctions between the von Neumann and L\"{u}ders
pictures from the viewpoint of state decoherence induced by the
measurement. All the aforementioned quantifiers could be utilized
to give the pair $\rmc^{(\clb)}$ and $\rmc^{(\clp)}$. To compare
various quantifiers of coherence, we consider the following
example.

Let us take a system consisting of two qubits. The $z$-component
of the total spin is represented by the operator
\begin{equation}
\asx_{z}=\bsg_{z}\otimes\pen_{2}+\pen_{2}\otimes\bsg_{z}
\, , \label{asxz}
\end{equation}
where $\bsg_{z}$ is a Pauli operators and $\pen_{2}$ is the
two-dimensional identity matrix. We clearly have
$\spc(\asx_{z})=\{2,0,0,-2\}$, so that the eigenvalue $0$ has
multiplicity $2$. In the case of L\"{u}ders-type measurement, we
deal with the three projectors
$|z_{0}z_{0}\rangle\langle{z}_{0}z_{0}|$,
$|z_{0}z_{1}\rangle\langle{z}_{0}z_{1}|+|z_{1}z_{0}\rangle\langle{z}_{1}z_{0}|$,
and $|z_{1}z_{1}\rangle\langle{z}_{1}z_{1}|$, where the kets are
such that
\begin{equation}
\bsg_{z}\,|z_{j}\rangle=(-1)^{j}\,|z_{j}\rangle
\, \label{zndef}
\end{equation}
for $j=0,1$. Following \cite{mayato12}, the refinement will be
taken as
$\asx_{z}+\bigl(\asx_{x}^{2}+\asx_{y}^{2}+\asx_{z}^{2}\bigr)/2$.
It has the spectrum $\{6,4,2,0\}$ with the corresponding
eigenvectors
\begin{equation}
|z_{0}z_{0}\rangle
\, , \qquad
|zz_{+}\rangle=\frac{|z_{0}z_{1}\rangle+|z_{1}z_{0}\rangle}{\sqrt{2}}
\ , \qquad
|z_{1}z_{1}\rangle
\, , \qquad
|zz_{-}\rangle=\frac{|z_{0}z_{1}\rangle-|z_{1}z_{0}\rangle}{\sqrt{2}}
\ . \label{eigs6420}
\end{equation}
After obtaining the value of the above refinement, we uniquely
reconstruct the value of $\asx_{z}$ \cite{mayato12}. The
pre-measurement state $\bro$ is transformed according to the
formulas
\begin{align}
\Phi_{\clb}(\bro)
&=\tr(\pim_{00}\bro)\,\pim_{00}+\tr(\pim_{+}\bro)\,\pim_{+}+\tr(\pim_{11}\bro)\,\pim_{11}+\tr(\pim_{-}\bro)\,\pim_{-}
\, , \label{vnbb}\\
\Phi_{\clp}(\bro)
&=\tr(\pim_{00}\bro)\,\pim_{00}+(\pim_{+}+\pim_{-})\bro\,(\pim_{+}+\pim_{-})+\tr(\pim_{11}\bro)\,\pim_{11}
\, . \label{vnpp}
\end{align}
Here, we denote one-rank projectors as
$\pim_{00}=|z_{0}z_{0}\rangle\langle{z}_{0}z_{0}|$,
$\pim_{11}=|z_{1}z_{1}\rangle\langle{z}_{1}z_{1}|$, and
$\pim_{\pm}=|zz_{\pm}\rangle\langle{z}z_{\pm}|$.

Effectively, distinctions between coherence decreasing with respect
to the von Neumann and L\"{u}ders pictures are brightly illuminated
in the two-dimensional subspace
${\mathrm{span}}\bigl\{|z_{0}z_{1}\rangle,|z_{1}z_{0}\rangle\bigr\}$.
Hence, we will mainly focus on two-dimensional matrices supported on
this subspace. It is obvious that such density matrices are
invariant under the action of (\ref{vnpp}), so that
$\Delta\rmc(\bro):=\rmc^{(\clb)}(\bro)$. With respect to the basis
$\bigl\{|zz_{+}\rangle,|zz_{-}\rangle\bigr\}$, we write
\begin{equation}
\bro=
\begin{pmatrix}
u & v^{*} \\
v & 1-u
\end{pmatrix}
, \label{sqdm}
\end{equation}
where real $u\in[0;1]$. The eigenvalues
$\lambda_{\pm}=1/2\pm\sqrt{1/4-{\mathrm{det}}(\bro)}$ satisfy
$0\leq\lambda_{\pm}\leq1$, whence $|v|\leq\sqrt{u(1-u)}\,$. It is
obvious here that
\begin{equation}
\Delta\rmc_{\ell_{1}}(\bro)=2|v|
\, . \label{dell1}
\end{equation}
As explicitly said in \cite{robcoh16}, for a qubit the robustness
of coherence is equal to the doubled modulus of an off-diagonal
element, so that $\Delta\rmr(\bro)=2|v|$. We also get
$\Delta\rmw(\bro)=2|v|$, whenever both the elements $u$ and $1-u$
are not less than $|v|$. In the situation considered, the three
measures gives the same term $2|v|$, which characterizes the level
of residual coherence in the L\"{u}ders picture. However, this is
not the case for a more general situation.

Let us consider coherence quantifiers based on the relative
$\alpha$-entropies. For some values of $\alpha$, we can write
relatively simple expressions:
\begin{align}
\Delta\rmc_{1}(\bro)&=h_{1}(u)-h_{1}(\lambda_{+})
\, , \label{delc1}\\
\Delta\rmc_{2}(\bro)&=
\left(
\sqrt{u^{2}+|v|^{2}}+
\sqrt{(1-u)^{2}+|v|^{2}}
\,\right)^{\!2}-1
\, , \label{delc2}
\end{align}
where $h_{1}(u):=-\,u\ln{u}-(1-u)\ln(1-u)$ is the binary Shannon
entropy. Of course, these values are different. It is interesting
that they are maximized for the same pure state, which can be
expressed as
$\sqrt{u}\,|zz_{+}\rangle+\sqrt{1-u}\,\exp(\iu\varphi)\,|zz_{-}\rangle$,
where $\varphi$ is the argument of $v$. The latter also maximizes
the term $\Delta\rmc_{\ell_{1}}(\bro)=\Delta\rmr(\bro)=2|v|$. In
general, different approaches to quantification of the level of residual
coherence lead to similar conclusions.

\section{On characteristics of coherence decreases in POVM-measurements}\label{sec4}

In this section, we address the question of how to describe the
decrease of a coherence in generalized quantum measurements. The
initial way to approach the notion of coherence is to represent
quantum states with respect to an orthonormal basis. We have already
seen that an extension to projective measurements is sufficiently
immediate. It is well known that any POVM-measurement can be
considered as a projective one in suitably extended space. In
principle, this possibility is established by the Naimark theorem. A
detailed description of general construction can be found, e.g. in
section 2.3.2 of \cite{watrous1}. We will restrict a consideration
to the case of rank-one POVMs, which is especially important for
several reasons. Due to the results of \cite{davies78}, for many
tasks the optimal POVM can be built of rank-one elements. Overall,
the method of constructing a projective measurement is sketched as
follows (see, e.g. section 3.1 of \cite{preskill}). Let
$\bigl\{|\mu_{j}\rangle\bigl\}_{j=1}^{N}$ be a set of sub-normalized
vectors that form a rank-one POVM with elements
\begin{equation}
\mm_{j}=|\mu_{j}\rangle\langle\mu_{j}|
\, . \label{mumun}
\end{equation}
By $\mu_{ij}$, we will mean $i$-th component of $j$-th vector
$|\mu_{j}\rangle$ with respect to the calculation basis. Due to
(\ref{comrn}), $d$ rows of the $d\times{N}$-matrix $[[\mu_{ij}]]$
are mutually orthogonal. By adding $(N-d)$ new rows, this matrix
can be converted into a unitary $N\times{N}$-matrix. Its columns
denoted by $|\wmu_{j}\rangle$ form an orthonormal basis $\wclb$ in
the corresponding $N$-dimensional space. As a block matrix, each
column is now written as
\begin{equation}
|\wmu_{j}\rangle:=
\begin{pmatrix}
 |\mu_{j}\rangle \\
 |\mu_{j}^{\prime}\rangle
\end{pmatrix}
 . \label{wum}
\end{equation}
As a result, we obtain some orthonormal and complete set of
vectors in the space $\widetilde{\hh}=\hh\oplus\hh^{\prime}$. In
general, there is more than one ways to build such orthonormal
basis, since one has a freedom to rotate vectors of the
ancillary space $\hh^{\prime}$ unitarily. The original density matrix is
rewritten as $\wbro=\dig(\bro,\zmx)$, so that for $\alpha>0$ we
get
\begin{equation}
\langle\wmu_{i}|\wbro^{\alpha}|\wmu_{j}\rangle=\langle\mu_{i}|\bro^{\alpha}|\mu_{j}\rangle
\, . \label{mawma}
\end{equation}
We also note that the above unitary freedom does not alter matrix
elements of the form (\ref{mawma}). Using the constructed
orthonormal basis $\wclb$, we are ready to put the set of
incoherent states and, herewith, to manage various coherence
quantifiers. Due to (\ref{mawma}), the $\ell_{1}$-norm of
coherence and relative-entropy-based quantifiers are expressed
immediately through the original terms related solely to $\hh$. In
particular, we write
\begin{align}
\rmc_{\ell_{1}}^{(\wclb)}(\wbro)
&=
\sum\nolimits_{i\neq{j}}\bigl|\langle\wmu_{i}|\wbro|\wmu_{j}\rangle\bigr|
=\sum\nolimits_{i\neq{j}}\bigl|\langle\mu_{i}|\bro|\mu_{j}\rangle\bigr|
\, , \label{n1wclb}\\
\rmc_{1}^{(\wclb)}(\wbro)
&=
H_{1}(p_{j})-\rms_{1}(\wbro)=H_{1}(p_{j})-\rms_{1}(\bro)
\, , \label{c1wclb}\\
\rmc_{\alpha}^{(\wclb)}(\wbro)
&=
\frac{1}{\alpha-1}
\left\{
\Bigl(
\sum\nolimits_{j}\langle\mu_{j}|\bro^{\alpha}|\mu_{j}\rangle^{1/\alpha}
\Bigr)^{\!\alpha}
-1
\right\}
 . \label{cawclb}
\end{align}
In (\ref{c1wclb}), we take into account that
$p_{j}=\langle\mu_{j}|\bro|\mu_{j}\rangle$ and the matrices
$\wbro$ and $\bro$ have the same non-zero eigenvalues. In
(\ref{cawclb}), we merely used (\ref{mawma}). We see that the
coherence quantifiers (\ref{n1wclb})--(\ref{cawclb}) are certainly
independent of the aforementioned unitary freedom. Due to this
fact, we will further focus just on such quantifiers. Immediately following
the measurement, one deals with a state completely incoherent
with respect to $\wclb$. Thus, any chosen quantifier can be used
to characterize the degree of coherence losses during the
measurement.

To exemplify the above approach, we apply it to the POVM-measurement
designed for unambiguous state discrimination. There exist two basic
approaches to discriminate between non-identical pure states
\begin{equation}
|\theta_{+}\rangle=
\begin{pmatrix}
\cos\theta \\
\sin\theta
\end{pmatrix}
 , \qquad
|\theta_{-}\rangle=
\begin{pmatrix}
\cos\theta \\
-\sin\theta
\end{pmatrix}
 . \label{etst}
\end{equation}
The Helstrom scheme optimizes the average probability of correct
answer. The second approach is known as unambiguous
discrimination. It sometimes gives an inconclusive answer, but never
makes an error of mis-identification. Of course, the measurement is
designed to minimize a fraction of inconclusive outcomes. There are
disputable questions connected with applications of unambiguous
discrimination in an individual attack on protocols of quantum
cryptography \cite{brandt05,shapiro06,rastfpb}.

By $\eta=\cos2\theta$, we denote the inner product, and restrict 
consideration to $\theta\in(0;\pi/2)$ -- that is, to non-identical
and non-orthogonal states. The POVM elements $\mm_{\pm}$ and
$\mm_{?}$ are expressed according to (\ref{mumun}) in terms of
sub-normalized vectors
\begin{equation}
|\mu_{\pm}\rangle=\frac{1}{\sqrt{1+\eta}}
\begin{pmatrix}
\sin\theta \\
\pm\cos\theta
\end{pmatrix}
 , \qquad
|\mu_{?}\rangle=\sqrt{\frac{2\eta}{1+\eta}}
\begin{pmatrix}
1 \\
0
\end{pmatrix}
 . \label{muth}
\end{equation}
After building a unitary $3\times3$-matrix, we obtain the
corresponding orthonormal basis $\wclb$ with vectors
\begin{equation}
|\wmu_{\pm}\rangle=\frac{1}{\sqrt{1+\eta}}
\begin{pmatrix}
\sin\theta \\
\pm\cos\theta \\
\sqrt{\eta}\,e^{\iu\gamma}
\end{pmatrix}
 , \qquad
|\wmu_{?}\rangle=\frac{1}{\sqrt{1+\eta}}
\begin{pmatrix}
\sqrt{2\eta} \\
0 \\
-\,\sqrt{1-\eta}\,e^{\iu\gamma}
\end{pmatrix}
 . \label{wmuth}
\end{equation}
Here, the phase factor $e^{\iu\gamma}$ reflects a unitary freedom in
the ancillary one-dimensional space. For the given probability
distribution $\{p_{+},p_{-},p_{?}\}$, the coherence measure
(\ref{c1wclb}) is maximal for pure states. It is instructive to
begin studies of coherence losses during the measurement with a pure
state. We further focus on the relative-entropy-based quantifiers.

What effect would the POVM-measurement have on a general initial
state, and are the states $|\theta_{\pm}\rangle$ special? With
respect to the calculation basis, we write $|\psi\rangle\in\hh$
and $|\wsi\rangle\in\widetilde{\hh}$ in the form
\begin{equation}
|\psi\rangle=
\begin{pmatrix}
\cos\vartheta \\
 e^{\iu\varphi}\sin\vartheta
\end{pmatrix}
 , \qquad
|\wsi\rangle=
\begin{pmatrix}
\cos\vartheta \\
 e^{\iu\varphi}\sin\vartheta \\
 0
\end{pmatrix}
\ . \label{psis}
\end{equation}
Assuming $\varphi\in[0;2\pi]$, we restrict consideration to the
values $\vartheta\in[0;\pi/2]$. Calculating inner products of the
form $\langle\mu_{j}|\psi\rangle$, we obtain the following
expressions of the chosen coherence quantifiers:
\begin{align}
\rmc_{1}^{(\wclb)}(|\wsi\rangle)
&=-\,p_{+}\ln{p}_{+}-p_{-}\ln{p}_{-}-p_{?}\ln{p}_{?}
\, , \label{c1pur}\\
\rmc_{\alpha}^{(\wclb)}(|\wsi\rangle)
&=
\frac{1}{\alpha-1}
\left\{
\left[
p_{+}^{1/\alpha}+p_{-}^{1/\alpha}+p_{?}^{1/\alpha}
\right]^{\alpha}
-1
\right\}
 , \label{capur}
\end{align}
where the probabilities are expressed as
\begin{align}
p_{+}&=\frac{\sin^{2}(\theta+\vartheta)-\sin2\theta\sin2\vartheta\sin^{2}\varphi/2}{1+\eta}
\ , \label{pplus}\\
p_{-}&=\frac{\sin^{2}(\theta+\vartheta)-\sin2\theta\sin2\vartheta\cos^{2}\varphi/2}{1+\eta}
\ , \label{pminus}\\
p_{?}&=\frac{2\eta\cos^{2}\vartheta}{1+\eta}
\ . \label{pinc}
\end{align}
It can be shown that the right-hand sides of (\ref{c1pur}) and
(\ref{capur}) are concave with respect to probability
distributions. This property should not be confused with
(\ref{conmixa}), since the above formulas are restricted to pure
states solely. In effect, we can rewrite (\ref{capur}) as
\begin{equation}
\rmc_{\alpha}^{(\wclb)}(|\wsi\rangle)=
\frac{\|p\|_{1/\alpha}-1}{\alpha-1}
\ , \label{nlal}
\end{equation}
where, for $\beta>0$, the norm-like function is defined as
$\|p\|_{\beta}:=\left(\sum\nolimits_{j}
p_{j}^{\beta}\right)^{1/\beta}$. Then the above-mentioned concavity
directly follows from the Minkowski inequality. We refrain from
presenting the details here. For each of the quantifiers
(\ref{c1pur}) and (\ref{capur}), one aims to find the minimal and
maximal values at the given $\theta$.

Let $\vartheta$ be fixed; then the terms $p_{?}$ and
$p_{+}+p_{-}=1-p_{?}$ are fixed as well. Inspecting the
corresponding derivative, we have arrived at a conclusion. Varying
$\varphi$ at the fixed $\vartheta$, the quantifier (\ref{capur})
is maximized for $p_{+}=p_{-}$, when
$\sin^{2}\varphi/2=\cos^{2}\varphi/2$. Hence, the relative phase
in (\ref{psis}) is equal to $\pm\pi/2$. The value of the coherence
quantifier is then expressed by
\begin{equation}
\frac{1}{\alpha-1}\left\{\left[
2p_{+}^{1/\alpha}+(1-2p_{+})^{1/\alpha}
\right]^{\alpha}
-1\right\}
. \label{capurax}
\end{equation}
In addition, the quantifier (\ref{capur}) is
minimized, when the distinction between $p_{+}$ and $p_{-}$ is made
as large as possible. We should further optimize the obtained
expressions by varying $\vartheta$. Concerning the maximum, this
task is realized through usual calculus.

Let us inspect the derivative of (\ref{capurax}) with respect to
$p_{+}$. It vanishes for $p_{+}=1-2p_{+}$, whence
$p_{+}=p_{-}=p_{?}=1/3$. Substituting the latter into
(\ref{capurax}) finally gives
\begin{equation}
\max\rmc_{\alpha}^{(\wclb)}(|\wsi\rangle)=\frac{3^{\alpha-1}-1}{\alpha-1}
=-\ln_{\alpha}\!\left(\frac{1}{3}\right)
 . \label{mln1p3}
\end{equation}
Here, the $\alpha$-logarithm is given by
$\ln_{\alpha}(z)=\bigl(z^{1-\alpha}-1\bigr)/(1-\alpha)$ for
$0<\alpha\neq1$ and real $z>0$. Due to (\ref{pinc}), the equality
$p_{?}=1/3$ is possible only for $2\eta/(1+\eta)\geq1/3$, whence
$\eta\geq1/5$. For $\eta<1/5$, the right-hand side of (\ref{mln1p3})
cannot be reached. The inequality $p_{?}<1/3$ leads to
$p_{+}=p_{-}>1/3$ and negative values of the derivative. So, the
function (\ref{capurax}) decreases with growth of $p_{+}>1/3$. To
maximize it, we should make $p_{?}$ as large as possible. Taking
$\cos^{2}\vartheta=1$, one gets
\begin{equation}
\max\rmc_{\alpha}^{(\wclb)}(|\wsi\rangle)=
\frac{1}{\alpha-1}\left\{\left[
2p_{+}^{1/\alpha}+p_{?}^{1/\alpha}
\right]^{\alpha}
-1\right\}
\ , \qquad
p_{+}=p_{-}=\frac{1-\eta}{2(1+\eta)}
\ , \label{mapurax}
\end{equation}
$p_{?}=2\eta/(1+\eta)$. This expression of the maximum holds for
$0<\eta<1/5$. The maximizing states are such that only the first
component is non-zero. When $\eta\geq1/5$, the maximizing states are
expressed as
\begin{equation}
|\psi_{\max}\rangle=
\begin{pmatrix}
\cos\vartheta \\
 \pm\iu\sin\vartheta
\end{pmatrix}
\ , \qquad
\cos^{2}\vartheta=\frac{1+\eta}{6\eta}
\ . \label{psim}
\end{equation}
In this interval of values of $\eta$, the relative phase of two
components of the maximizing state should be equal to $\pm\pi/2$.
For all $\alpha>0$, the coherence $\alpha$-quantifier
$\rmc_{\alpha}^{(\wclb)}\bigl(|\wsi\rangle\bigr)$ is maximized by
the same states of the principal space $\hh$.

In general, exact analytical expressions of the
minimum for arbitrary $\alpha>0$ are difficult to obtain. These
difficulties originate in the structure of the domain, in which
quantifiers should be minimized. In the three-dimensional real
space, the conditions $p_{\pm}\geq0$, $p_{?}\geq0$ and
$p_{+}+p_{-}+p_{?}=1$ specify the triangle with the
vertices $(1,0,0)$, $(0,1,0)$, and $(0,0,1)$. It must be stressed
that the three probabilities are connected by the relations
(\ref{pplus})--(\ref{pinc}). Combining (\ref{pplus}) with
(\ref{pminus}), one gets
\begin{equation}
|p_{+}-p_{-}|=\frac{\sin2\theta\sin2\vartheta\>|\cos\varphi|}{1+\eta}
\leq\frac{\sin2\theta\sin2\vartheta}{1+\eta}
\ . \label{difpp}
\end{equation}
With respect to the rotated coordinate system with coordinates
$x=(p_{+}+p_{-})/\sqrt{2}$, $y=(-\,p_{+}+p_{-})/\sqrt{2}$, and
$z=p_{?}$, the inequality (\ref{difpp}) fixes an elliptic solid
cylinder with the surface
\begin{equation}
\frac{y^{2}}{a^{2}}+\frac{(z-b)^{2}}{b^{2}}=1
\, , \qquad
a=\sqrt{\frac{1-\eta}{2(1+\eta)}}
\ , \qquad
b=\frac{\eta}{1+\eta}
\ . \label{elleq}
\end{equation}
Cutting the above cylinder in the plane $p_{+}+p_{-}+p_{?}=1$, we
get the domain of allowed values of the three probabilities. The
domain boundary is an ellipse inscribed in the triangle with the
vertices $(1,0,0)$, $(0,1,0)$, and $(0,0,1)$. It touches the three
sides in the points $(1-\eta,0,\eta)$, $(0,1-\eta,\eta)$, and
$(1/2,1/2,0)$. Note that the two touching points correspond to the
states $|\theta_{+}\rangle$ and $|\theta_{-}\rangle$
respectively. The minimum of the concave function (\ref{capur})
relative to a convex set is attained at one of its extreme points
(see, e.g., corollary 32.3.2 of \cite{rockaffellar}). Hence, this
quantifier should be minimized with respect to the
elliptic boundary of the domain. As general closed formulas are
difficult to express, we visualize the results for especially
interesting choices of $\alpha>0$.

\begin{figure}
\centering
\includegraphics[width=8.0cm]{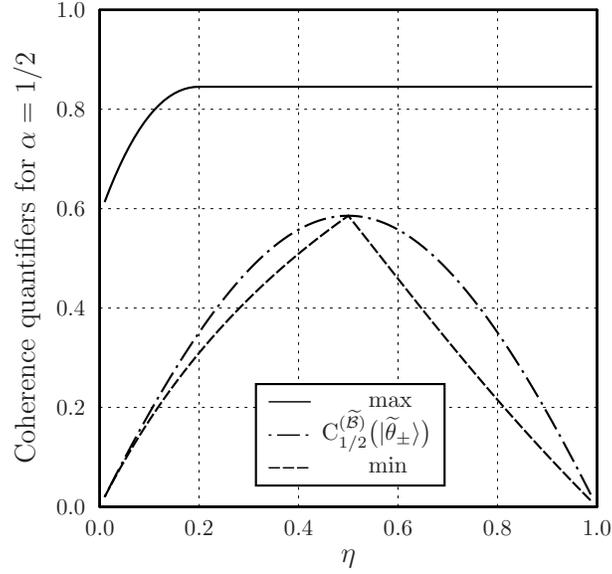}
\caption{\label{qa12} Coherence $\alpha$-quantifiers for
$\alpha=1/2$ versus $\eta\in(0;1)$.}
\end{figure}

We begin with the case $\alpha=1/2$, in which sufficiently
simple expressions take place. The corresponding quantifier is
expressed as
\begin{equation}
\rmc_{1/2}^{(\wclb)}(|\wsi\rangle)=
2-2\sqrt{p_{+}^{2}+p_{-}^{2}+p_{?}^{2}}
\, . \label{misqu}
\end{equation}
To minimize (\ref{misqu}), we should maximize the sum of squares
of the three probabilities. By the usual algebra, one gets
\begin{equation}
p_{+}^{2}+p_{-}^{2}+p_{?}^{2}=\frac{1}{1+2\eta}+
\frac{4\eta^{2}-1}{2\eta^{2}}
\left(
p_{?}-\frac{1}{1+2\eta}
\right)^{\!2}
 . \label{sumsqs}
\end{equation}
So, the result depends on the sign of the factor $4\eta^{2}-1$,
where $\eta\in(0;1)$. Combining (\ref{misqu}) with (\ref{sumsqs})
finally gives the answer written as
\begin{equation}
\min\rmc_{1/2}^{(\wclb)}(|\wsi\rangle)=
\begin{cases}
 2-\frac{2}{\sqrt{1+2\eta}}\ , & \text{for}\ 0<\eta\leq1/2\, , \\
 2-2\left[\frac{1}{1+2\eta}+\frac{(4\eta^{2}-1)(1+3\eta)^{2}}{2(1+\eta)^{2}(1+2\eta)^{2}}\right]^{1/2} , & \text{for}\ 1/2\leq\eta<1\, .
\end{cases}
\label{min12}
\end{equation}
It is instructive to compare (\ref{min12}) with the quantity
\begin{equation}
\rmc_{1/2}^{(\wclb)}(|\widetilde{\theta}_{\pm}\rangle)=
2-2\sqrt{(1-\eta)^{2}+\eta^{2}}
\, . \label{thet12}
\end{equation}
In figure \ref{qa12}, we draw the maximal and minimal values of
$\rmc_{1/2}^{(\wclb)}\bigl(|\wsi\rangle\bigr)$ together with
(\ref{thet12}) as functions of the parameter $\eta$. Although the
$1/2$-quantifier is not minimized exactly by
$|\widetilde{\theta}_{\pm}\rangle$, these states give almost
minimal values.

\begin{figure}
\centering
\includegraphics[width=8.0cm]{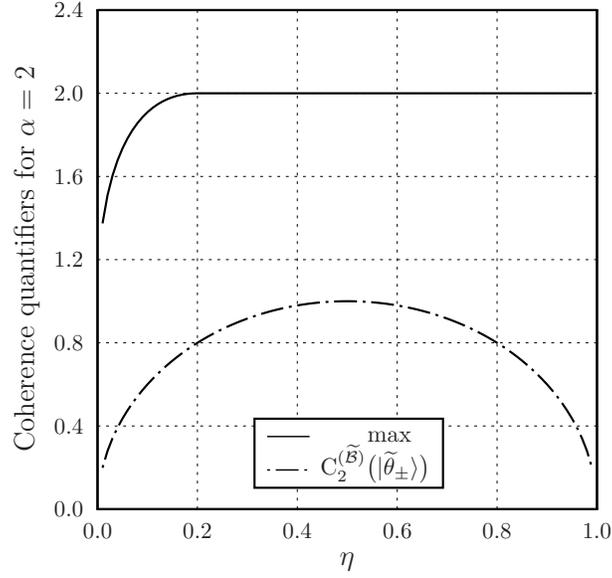}
\caption{\label{qa2} Coherence $\alpha$-quantifiers for $\alpha=2$
versus $\eta\in(0;1)$. The curve
$\rmc_{2}^{(\wclb)}(|\widetilde{\theta}_{\pm}\rangle)$
also gives the minimal value.}
\end{figure}

The value $\alpha=2$ leads to another relatively simple choice. It
turns out that the $2$-quantifier coincides here with the
$\ell_{1}$-norm of coherence. For a pure state, the logarithmic
coherence of \cite{apwl17} can be interpreted in terms of the
R\'{e}nyi entropy of certain order. Our approach leads to another
entropy-based reformulation of the $\ell_{1}$-norm of coherence.
In the case of pure states, one has
\begin{align}
&\rmc_{2}^{(\wclb)}(|\wsi\rangle)=
(\sqrt{p_{+}}+\sqrt{p_{-}}+\sqrt{p_{?}})^{2}-1
\nonumber\\
&=2\sqrt{p_{+}p_{-}}+2\sqrt{p_{-}p_{?}}+2\sqrt{p_{?}p_{+}}
=\rmc_{\ell_{1}}^{(\wclb)}(|\wsi\rangle)
\, . \label{qael1}
\end{align}
It follows from $p_{+}+p_{-}=1-z$ and (\ref{elleq}) that
\begin{equation}
2\sqrt{p_{+}p_{-}}=\sqrt{(1-z)^{2}+2y^{2}}=\frac{|\eta-z|}{\eta}
 . \label{key2}
\end{equation}
Taking $z\in[0;2b]$, we wish to minimize the sum of square roots
of the three probabilities. Due to (\ref{key2}), this sum appears
as
\begin{equation}
f(z)=\sqrt{p_{+}}+\sqrt{p_{-}}+\sqrt{p_{?}}=\sqrt{1-z+\eta^{-1}|\eta-z|}+\sqrt{z}
\, . \label{fzdf}
\end{equation}
For $0\leq{z}\leq\eta$, we deal with the concave function
$\sqrt{2-z/b}+\sqrt{z}$. Its minimal value is one of two least
values $f(0)=\sqrt{2}$ and $f(\eta)=\sqrt{1-\eta}+\sqrt{\eta}$.
Except for $\eta=1/2$, the term $f(\eta)$ is strictly less than
$f(0)$. For $\eta\leq{z}\leq2b$, our concave function is written
as
\begin{equation}
\sqrt{z}\,\sqrt{\frac{1-\eta}{\eta}}+\sqrt{z}
\,  . \label{sz}
\end{equation}
Here, we have $f(\eta)=\sqrt{1-\eta}+\sqrt{\eta}$ again and
$f(2b)=f(\eta)\sqrt{2/(1+\eta)}$. To sum up, we conclude that
\begin{equation}
\min\rmc_{2}^{(\wclb)}(|\wsi\rangle)=\sqrt{1-\eta}+\sqrt{\eta}
=\rmc_{2}^{(\wclb)}(|\widetilde{\theta}_{\pm}\rangle)
\, . \label{min2}
\end{equation}
That is, the states $|\widetilde{\theta}_{\pm}\rangle$ to be
discriminated minimize the coherence $2$-quantifier exactly. In view
of (\ref{qael1}), the same conclusion holds for the $\ell_{1}$-norm
of coherence. In figure \ref{qa2}, we show the maximal and minimal
values of $\rmc_{2}^{(\wclb)}(|\wsi\rangle)$ as functions of the
parameter $\eta$. Overall, the picture is similar to that is related
to the case $\alpha=1/2$. The only distinction is that the states
$|\widetilde{\theta}_{\pm}\rangle$ exactly minimize the
quantifier for $\alpha=2$.

\begin{figure}
\centering
\includegraphics[width=8.0cm]{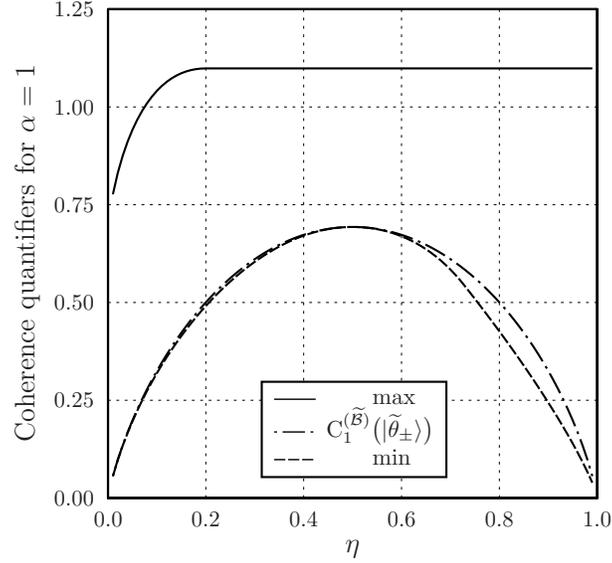}
\caption{\label{qa1} Coherence $\alpha$-quantifiers for $\alpha=1$
versus $\eta\in(0;1)$.}
\end{figure}

To complete the discussion, we also consider the value $\alpha=1$.
For pure states, the corresponding measure of coherence appears as
the Shannon entropy of generated probability distribution. For
$|\widetilde{\theta}_{\pm}\rangle$, we obtain the binary Shannon
entropy
\begin{equation}
\rmc_{1}^{(\wclb)}(|\widetilde{\theta}_{\pm}\rangle)=h_{1}(\eta)=
\!{}-(1-\eta)\ln(1-\eta)-\eta\ln\eta
\, . \label{thet1}
\end{equation}
The minimization is difficult to formulate analytically.
Nevertheless, we can present the results of numerical investigation.
In figure \ref{qa1}, we draw the maximal and minimal values of
$\rmc_{1}^{(\wclb)}\bigl(|\wsi\rangle\bigr)$ together with
(\ref{thet1}) as functions of the parameter $\eta$. Similarly to
figure \ref{qa12}, the states $|\widetilde{\theta}_{\pm}\rangle$ give
almost minimal values.

We have studied characteristics of coherence losses during
unambiguous state discrimination. Various coherence quantifiers were
actually connected with the extended space
$\widetilde{\hh}=\hh\oplus\hh^{\prime}$. On the other hand, the
states under consideration have non-zero components only in the
principal space $\hh$. For pure states, the maximum of the coherence
$\alpha$-quantifier as a function of $\eta$ is expressed by
(\ref{mln1p3}) and (\ref{mapurax}). To study minimal values, we
choose the $\alpha$-quantifiers for $\alpha=1/2,1,2$. Due to
(\ref{qael1}), our choice includes the $\ell_{1}$-norm of coherence
as well. The measurement for unambiguous state discrimination is
designed to distinguish states $|\theta_{+}\rangle$ and
$|\theta_{-}\rangle$ without the error of mis-identification. For
these states, visible losses of quantum coherence are minimal or
almost minimal.

\section{Conclusions}\label{sec5}

We have considered some coherence quantifiers from the viewpoint of
their changes in quantum measurements. For an observable with
possibly degenerate eigenvalues, there exist two different ways to
formulate the state immediately following a measurement. These ways are
commonly referred as the von Neumann and L\"{u}ders reduction rules.
The latter implies the quantum operation written in terms of the
corresponding projectors. Due to another choice of incoherent
states, coherence quantifiers are defined via optimization over a
larger set of allowed states. We applied this approach to quantities
based on quantum $\alpha$-divergences of the Tsallis type. It was
shown that such coherence quantifiers succeed the same formal
properties as defined with respect to an orthonormal bases. The
robustness of coherence and the coherence weight have also been addressed briefly.
To illustrate distinctions between the L\"{u}ders
and von Neumann pictures in the sense of coherence losses, we
considered an example of some spin observable with a degenerate
eigenvalue. Different coherence measures lead to similar conclusions
about the level of residual coherence.

Another interesting question concerns ways to characterize decreases
of quantum coherence in POVM measurements. We focused on rank-one
POVMs, since they just include principal features of the problem. In
this case, we finally deal with some orthonormal basis in the
extended space. Hence, basic ways to quantifying the amount of
quantum coherence can be applied. Of course, the construction described
contains a unitary freedom. Since the $\ell_{1}$-norm
of coherence and coherence $\alpha$-quantifiers are expressed via
matrix elements of the density matrix and its powers, they are
independent of this freedom. The proposed approach is exemplified
using a POVM designed for unambiguous discrimination of two
non-orthogonal pure states. It can naturally be converted into
orthonormal basis in the three-dimensional space. Taking arbitrary
pure state, we study the maximal and minimal values of the chosen
quantifiers as function of the overlap between two states to be
identified. In the sense of coherence losses, these two states
clearly reveal some extreme properties.

\end{document}